\documentclass[twocolumn, tighten]{aastex701}

\begin{document}

\title{The Shape of (486958) Arrokoth}

\author[0000-0003-0333-6055]{Simon B. Porter}
\affiliation{Southwest Research Institute, 1301 Walnut St, Suite 400, Boulder, CO 80302, USA}
\email[show]{simon.porter@swri.org}

\author[0000-0003-3045-8445]{Kelsi N. Singer}
\affiliation{Southwest Research Institute, 1301 Walnut St, Suite 400, Boulder, CO 80302, USA}
\email{kelsi.singer@swri.org}

\author[0000-0003-1316-5667]{Paul M. Schenk}
\affiliation{Lunar and Planetary Institute, 3600 Bay Area Blvd., Houston, TX 77058, USA}
\email{schenk@lpi.usra.edu}

\author[0000-0002-3323-9304]{Anne J. Verbiscer}
\affiliation{Department of Astronomy, University of Virginia, P.O. Box 400325, Charlottesville, VA, 22904, USA}
\email{av4n@virginia.edu}

\author[0000-0001-8821-5927]{Susan D. Benecchi}
\affiliation{Planetary Science Institute, 1700 East Fort Lowell, Suite 106, Tucson, AZ, 85719, USA}
\email{susank@psi.edu}

\author[0000-0003-4452-8109]{John R. Spencer}
\affiliation{Southwest Research Institute, 1301 Walnut St, Suite 400, Boulder, CO 80302, USA}
\email{john.spencer@swri.org}

\author[0000-0002-3672-0603]{Joel Wm. Parker}
\affiliation{Southwest Research Institute, 1301 Walnut St, Suite 400, Boulder, CO 80302, USA}
\email{joel.parker@swri.org}

\author[0000-0002-4644-0306]{Pontus Brandt}
\affiliation{Johns Hopkins University Applied Physics Laboratory, Laurel, MD, USA}
\email{pontus.brandt@jhuapl.edu}

\author{S. Alan Stern}
\affiliation{Southwest Research Institute, 1301 Walnut St, Suite 400, Boulder, CO 80302, USA}
\email{alan.stern@swri.org}

\begin{abstract}
Here we present an updated shape model of (486958) Arrokoth, the bilobate Kuiper Belt Object (KBO) which the NASA
New Horizons spacecraft flew past in 2019. 
This updated shape model uses all of the resolved images of Arrokoth obtained by the New Horizons LOng Range Reconnaissance Imager (LORRI).
We developed an updated shape modeling algorithm which allowed the shape and rotational pole of Arrokoth to be fit
to much better quality with an efficient use of GPU-accelerated features.
The resulting model of Arrokoth's contact binary shape is significantly thicker and of larger volume than the one previously published immediately after the flyby
by Spencer et al (2020).
We show that Arrokoth's smaller lobe Weeyo is roughly spherical in shape, while the larger lobe Wenu is more flattened,
with the volume ratio between the lobes being roughly 2:1.
Owing to Wenu's oblate shape, Arrokoth's rotational lightcurve would have significantly lower mean reflectance when viewed from 
subobserver latitudes that would have shown lightcurve variation. We discuss the impact this may have on estimates of
the frequency of contact binaries in the Kuiper Belt.
We also discuss the implications of this shape for the formation of Arrokoth, 
particularly in the context of the Streaming Instability.
\end{abstract}

\section{Introduction}

On January 1, 2019, NASA’s New Horizons spacecraft flew about 3500 km from (486958) Arrokoth,
a Cold Classical Kuiper Belt Object (KBO), making it the farthest solar system object from the Sun to be studied up close 
by a spacecraft \citep{2019Sci...364.9771S}.
The Kuiper Belt was named after the predictions of \citet{1974CeMec...9..321K} of a large comet belt beyond Neptune, 
which had been first predicted by \citet{1949MNRAS.109..600E}, and was first confirmed to exist with the discovery of 1992 QB$_1$, now (15760) Albion 
\citep{1993Natur.362..730J}.
Albion is a roughly 100 km diameter object that has a mean distance from the Sun of about 44 AU in a near-circular
($e<0.1$), low inclination orbit \citep{2011Icar..213..693B}.
These characteristics made Albion not just the first small KBO to be discovered, but also the archetype for the specific class of 
``Cold Classical KBOs'' (CCKBOs) that appear to be in orbits which were not significantly perturbed during the giant planet
migration phase \citep{2005AJ....129.1117E}, thus making them effectively fossils of the early stages of planet formation 
\citep{2005Natur.435..459T}.
This lack of migration has allowed the CCKBOs to have suffered far fewer impacts than other populations beyond Neptune 
\citep{2019ApJ...872L...5G},
and the lack of impacts is proven out with the relatively high fraction of CCKBOs that are seen to be very wide (loosely-bound)
binaries \citep{2019Icar..334...62G}.
Indeed, the overall binary fraction for the CCKBOs is quite high, at 21\% detectable as binaries with the resolution to the 
Hubble Space Telescope (HST) Wide Field Camera 3 (WFC3) instrument 
\citep[1000 km;][]{Porter_2024},
and more tighter CCKBO binaries detected with an occultation \citep[350 km;][]{2020PSJ.....1...48L}
and two with direct imaging by the New Horizons spacecraft \citep[200 km;][]{2022PSJ.....3...46W}.
Ground-based lightcurve studies showed that at least half of CCKBOs could be contact binaries \citep{2019AJ....157..228T}, 
though lightcurves can only be used to detect the fraction of KBOs that are roughly equator-on as seen from Earth 
\citep{2021Icar..35614098S}. 

Even before New Horizons' flyby of Arrokoth, these discoveries led to the theory that all CCKBOs may have initially formed 
as separated binaries \citep{2017NatAs...1E..88F}, with many later either being split or merging into contact binaries.
A formation theory that supports the concept of (nearly) all CCKBOs forming as binary objects is the Streaming Instability 
\citep[SI;][]{2019NatAs...3..808N}.
The SI theory postulates that the primary mechanism that formed planetesimals in the outer proto-Solar disk was 
gas drag causing dust to clump into linear ``streams'', which then gravitationally collapse into solid bodies 
\citep[see][for far more details]{2007ApJ...662..613Y,2017ApJ...839...16C,2019ApJ...883..192A}.
A key constraint of SI formation is that it is fast, too fast to allow gas drag to bleed away much angular momentum from the 
collapsing dust clouds \citep{2007ApJ...662..613Y}.
The rapid gravitational collapse of the solid material while preserving angular momentum means that SI directly forms binary KBOs from
the disk \citep{2019NatAs...3..808N}, without any need for later binary capture \citep[e.g.][]{2002Natur.420..643G}.
Because of this rapid collapse, SI can even form triple or higher systems \citep{2019NatAs...3..808N}, 
though only one triple KBO is known to exist \citep[Lempo;][]{2010Icar..207..978B} 
with at least one more suspected from orbital analysis \citep[Altjira;][]{2025PSJ.....6...53N}.
Importantly, binaries formed from the SI process should have identical composition for both bodies, and show a preference
for prograde orbits, as this is the direction of the net angular momentum vector favored by gravitational shear of the
streams in the SI \citep{2019NatAs...3..808N}.
The similarity of compositions has been tested by both colors using HST \citep{2009Icar..200..292B}, and 
recently with the NIRSpec spectrometer on the James Webb Space Telescope (JWST) \citep{2024A&A...681L..17S}, 
with all studies finding a similarity of spectra between binary components.
\citet{2019Icar..334...62G} found that the known binary KBOs at the time indeed showed a prograde preference, and that preference is likely
preserved even for wide binary KBOs after billions of years of solar perturbations on the binary orbit \citep{2012Icar..220..947P}.
There is thus considerable evidence to support the idea that CCKBOs were primarily formed through the SI, and
primarily born as binaries, and that is the context we will use to analyze the shape and origin of Arrokoth.

\startlongtable
\begin{deluxetable*}{ccccccc}
\tablecaption{Description of LORRI image sequences used in this analysis.
\label{tab:im}}
\tablehead{  \colhead{} & \colhead{Observation} & \colhead{Exposure} & \colhead{Num.}
& \colhead{Distance} & \colhead{Solar} & \colhead{m/pixel} \\
\colhead{REQID\tablenotemark{a}} & \colhead{Midtime} & \colhead{Time (s)} & \colhead{Img.\tablenotemark{b}}
& \colhead{(km)\tablenotemark{c}} & \colhead{Phase\tablenotemark{d}} & \colhead{Res.\tablenotemark{e}} }
\startdata
\texttt{APROTNAV\_L1\_2018365A} & 2018-12-31 18:13:13.719 & 0.1506 & 6 & 589111 & 11.701$^\circ$ & 2970.3 \\
\texttt{APROTNAV\_L1\_2018365B} & 2018-12-31 18:46:13.719 & 0.1506 & 6 & 560529 & 11.699$^\circ$ & 2826.2 \\
\texttt{APROTNAV\_L1\_2018365C} & 2018-12-31 19:23:13.719 & 0.1506 & 6 & 528483 & 11.697$^\circ$ & 2664.6 \\
\texttt{APROTNAV\_L1\_2018365D} & 2018-12-31 20:00:13.719 & 0.1506 & 6 & 496436 & 11.695$^\circ$ & 2503.1 \\
\texttt{APROTNAV\_L1\_2018365E} & 2018-12-31 20:38:13.719 & 0.1506 & 6 & 463524 & 11.692$^\circ$ & 2337.1 \\
\texttt{APROTNAV\_L1\_2018365F} & 2018-12-31 21:28:13.719 & 0.1506 & 6 & 420218 & 11.687$^\circ$ & 2118.8 \\
\texttt{APROTNAV\_L1\_2018365G} & 2018-12-31 22:01:53.719 & 0.1506 & 6 & 391059 & 11.684$^\circ$ & 1971.7 \\
\texttt{APROTNAV\_L1\_2018365H} & 2018-12-31 22:46:13.719 & 0.1506 & 6 & 352662 & 11.679$^\circ$ & 1778.1 \\
\texttt{APROTNAV\_L1\_2018365I} & 2018-12-31 23:26:13.719 & 0.1506 & 6 & 318018 & 11.674$^\circ$ & 1603.5 \\
\texttt{APROTNAV\_L1\_2018365J} & 2018-12-31 23:46:28.719 & 0.1506 & 6 & 300480 & 11.671$^\circ$ & 1515.0 \\
\texttt{APROTNAV\_L1\_2019001A} & 2019-01-01 00:07:59.719 & 0.1506 & 6 & 281845 & 11.667$^\circ$ & 1421.1 \\
\texttt{APROTNAV\_L1\_2019001B} & 2019-01-01 00:29:30.719 & 0.1506 & 6 & 263210 & 11.663$^\circ$ & 1327.1 \\
\texttt{APROTNAV\_L1\_2019001C} & 2019-01-01 00:51:01.719 & 0.1506 & 6 & 244575 & 11.659$^\circ$ & 1233.2 \\
\texttt{APROTNAV\_L1\_2019001D} & 2019-01-01 01:12:32.719 & 0.1506 & 6 & 225941 & 11.655$^\circ$ & 1139.2 \\
\texttt{APROTNAV\_L1\_2019001E} & 2019-01-01 01:34:03.719 & 0.1506 & 6 & 207307 & 11.650$^\circ$ & 1045.3 \\
\texttt{APROTNAV\_L1\_2019001F} & 2019-01-01 01:55:34.719 & 0.1506 & 6 & 188674 & 11.645$^\circ$ & 951.3 \\
\texttt{APROTNAV\_L1\_2019001G} & 2019-01-01 02:17:05.719 & 0.1506 & 6 & 170041 & 11.640$^\circ$ & 857.4 \\
\texttt{APMAP\_L1\_2019001\_\_stare1} & 2019-01-01 02:45:44.553 & 0.1506 & 6 & 145235 & 11.635$^\circ$ & 732.3 \\
\texttt{APMAP\_L1\_2019001\_\_mosaic} & 2019-01-01 02:59:45.553 & 0.1506 & 4 & 133098 & 11.633$^\circ$ & 671.1 \\
\texttt{APMAP\_L1\_2019001\_\_stare2} & 2019-01-01 03:08:44.553 & 0.1506 & 6 & 125320 & 11.632$^\circ$ & 631.9 \\
\texttt{APRETARG\_L1\_2019001} & 2019-01-01 03:21:07.553 & 0.1506 & 12 & 114599 & 11.633$^\circ$ & 577.8 \\
\texttt{APROTNAV\_L1\_2019001I} & 2019-01-01 03:26:25.053 & 0.1506 & 6 & 110018 & 11.634$^\circ$ & 554.7 \\
\texttt{APROTNAV\_L1\_2019001J} & 2019-01-01 03:43:15.053 & 0.1506 & 6 & 95447 & 11.643$^\circ$ & 481.2 \\
\texttt{APROTNAV\_L1\_2019001K} & 2019-01-01 04:04:32.053 & 0.1506 & 6 & 77028 & 11.675$^\circ$ & 388.4 \\
\texttt{CA01-MAP\_L1\_2019001} & 2019-01-01 04:22:49.053 & 0.1506 & 49 & 61213 & 11.751$^\circ$ & 308.6 \\
\texttt{CA02-MAP\_L1\_2019001} & 2019-01-01 04:44:17.115 & 0.0256 & 7 & 42664 & 12.039$^\circ$ & 215.1 \\
\texttt{CA04-MAP\_L1\_2019001} & 2019-01-01 05:01:20.078 & 0.1006 & 31 & 27974 & 12.907$^\circ$ & 141.0 \\
\texttt{CA05-MAP\_L1\_2019001} & 2019-01-01 05:14:33.115 & 0.0256 & 7 & 16681 & 15.694$^\circ$ & 84.1 \\
\texttt{CA06-PREALLOC\_L1\_2019001} & 2019-01-01 05:26:54.615 & 0.0256 & 6 & 6622 & 32.585$^\circ$ & 33.4 \\
\texttt{CA07-LIMB\_L4\_2019001} & 2019-01-01 05:42:42.028 & 0.2006 & 13 & 8818 & 152.333$^\circ$ & 177.9 \\
\enddata
\tablenotetext{a}{Image sequence identifier; the prefix \texttt{KELR\_MU69\_} has been removed for clarity}
\tablenotetext{b}{Number of raw images used for the final stacked image}
\tablenotetext{c}{New Horizons-Arrokoth distance in kilometers}
\tablenotetext{d}{Sun-Arrokoth-New Horizons Solar Phase angle in degrees}
\tablenotetext{e}{Original image ground pixel scale in meters per pixel, not taking into account any smear}
\end{deluxetable*}

(486958) Arrokoth was provisionally designated 2014 MU$_{69}$ and also called ``PT1'' and ``Ultima Thule'' prior to formal naming
after the Powhatan word for Sky (in recognition of the traditional owners of the land on which the Johns Hopkins University
Applied Physics Laboratory sits). 
It was discovered in 2014 with a deep survey using HST/WFC3 \citep[for the full history of its discovery]{2024PSJ.....5..196B}.
Seen from Earth, Arrokoth appears very faint \citep[V$\sim$26;][]{2019Icar..334...11B} 
and is only reliably recoverable with HST for astrometry \citep{2018AJ....156...20P}.
Upon discovery it was immediately apparent from its orbital elements that Arrokoth was a CCKBO, and by early 2015, it was known to be the potential flyby target that required the smallest  
delta-v for targeting a New Horizons flyby. It was also clear that the flyby distance would have to be quite large
if the orbit was only constrained by HST astrometry \citep{2018AJ....156...20P}.
The New Horizons mission then initiated a campaign to observe multiple stellar occultations by Arrokoth to provide high-precision astrometric
points to constrain the KBO's orbit as well as measure its size, shape, and albedo \citep{2020AJ....159..130B}.
This campaign was uniquely enabled by access to the pre-release version of the ESA Gaia DR2 star catalog \citep{2018A&A...616A...1G}.
The first attempt to observe a stellar occultation by Arrokoth on 3 June 2017 using telescopes in northern Argentina and South Africa was not successful.
On 10 July 2017, the NASA/DLR Stratospheric Observatory for Infrared Astronomy (SOFIA) obtained a single, grazing occultation chord. 
From southern Argentina on 17 July 2017, the New Horizons occultation campaign finally successfully observed an Arrokoth occultation, obtaining five positive chords across the KBO
\citep{2020AJ....159..130B}.
About one year later in August 2018, the campaign observed a second multi-chord stellar occultation by Arrokoth in Senegal \citep{2020AJ....159..130B}.
Together, these two occultations (along with HST data and the Gaia stars) provided an orbit solution with  sufficiently high resolution 
that New Horizons was able to fly just 3536 km from Arrokoth and center the KBO within the full LORRI \citep[LOng Range Reconnaissance Imager;][]{2008SSRv..140..189C} field of view, thus revealing its shape in exquisite detail \citep{2019Sci...364.9771S}.
However, even before the flyby, the occultations (especially the multi-chord event in 2017) revealed that Arrokoth has a complex shape.
The 2017 occultation appeared to show two roughly circular objects, but it was not clear whether Arrokoth was a contact binary,
a separated binary, or just an elongated object with large terrain relief \citep{2020AJ....159..130B}.
The post-flyby knowledge that Arrokoth is indeed a contact binary is strong motivation for conducting future stellar occultation studies 
of CCKBO shapes, as will be discussed below.

The New Horizons spacecraft first acquired Arrokoth on August 16, 2018\footnote{https://pluto.jhuapl.edu/News-Center/News-Article.php?page=20180828}
with the LORRI 
 camera in ``$4\times4$'' photometric mode, 
but it showed no discernible lightcurve \citep{2020Sci...367.3999S}.
This was not surprising, as Arrokoth had shown no lightcurve to HST \citep{2019Icar..334...11B} 
and New Horizons was only looking 11$^\circ$ away from the Arrokoth-Earth direction.
It was not until December 30, 2018, 2 days before the closest approach, that LORRI (in ``$1\times1$'' high resolution imaging mode) was able to resolve 
the shape of Arrokoth, appearing to have a bowling pin-like profile\footnote{https://pluto.jhuapl.edu/News-Center/News-Article.php?page=20181231}.
While further imaging continued, additional images were not downlinked to Earth until after the flyby on January 1, 2019.
The first images returned after flyby (the ``\texttt{CA01}'' sequence) clearly showed the contact binary shape of Arrokoth.
In total, 30 image sequences were returned that showed the resolved shape of Arrokoth, in addition to one (``\texttt{CA07}'') that
showed the double crescent of Arrokoth as New Horizons looked back at it with LORRI in $4\times4$ mode, as shown in Table \ref{tab:im}.
Arrokoth was the first, and so far only, KBO to be imaged at sufficiently high resolution that its shape could be ascertained,
and that shape was a contact binary comprised of two lenticular lobes, with little apparent surface topography apart from
a very large crater (now named ``Sky'').
This is a shape unlike any other spacecraft target so far, 
but it is possibly indicative that a large fraction of KBOs formed by the Streaming Instability 
\citep{2020Sci...367.6620M, 2023PSJ.....4..176S}.

Preliminary versions of the shape model of Arrokoth published in \citet{2019Sci...364.9771S} and \citet{2020Sci...367.3999S},
both used a limited subset of the imagery and lacked variable albedo on the surface of Arrokoth.
This latter point is important, as Arrokoth does show considerable albedo variation (by about a factor of two) around the
interface between the lobes, in Sky Crater, and a few other features.
The shape fitting program used for that analysis has been significantly improved through the support of the New Horizons Second Kuiper Belt Extended Mission (KEM2), and here we present the results of those improvements.
In this work, we update the rotation period, shape, and pole solution of Arrokoth, and added albedo mapping of the imaged half of the body.
We present updated estimates of the volumes of the lobes, and map out the terrain relief on the imaged hemispheres of the
two lobes.
We also discuss the implications of the updated shape model for the formation of Arrokoth in the context of the SI,
as well as how the shape of Arrokoth affects estimates of how frequent Arrokoth-like contact binaries may be in the Cold Classical Kuiper Belt.
This work has been archived in the NASA Planetary Data System (PDS), and here we provide documentation,
context, and commentary on that product \citep{Stern_2025}.

\section{Shape Fitting}

\begin{figure*}
\plottwo{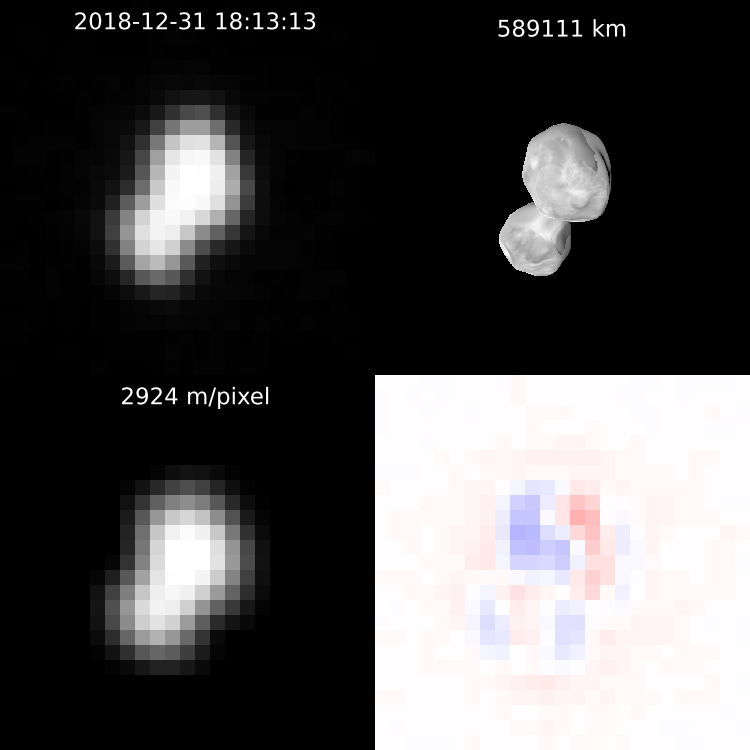}{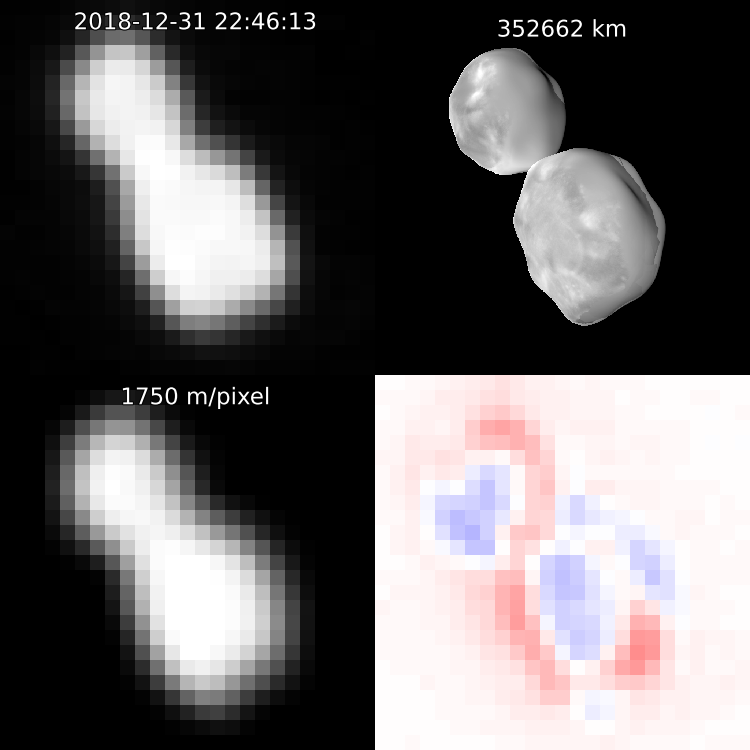}
\plottwo{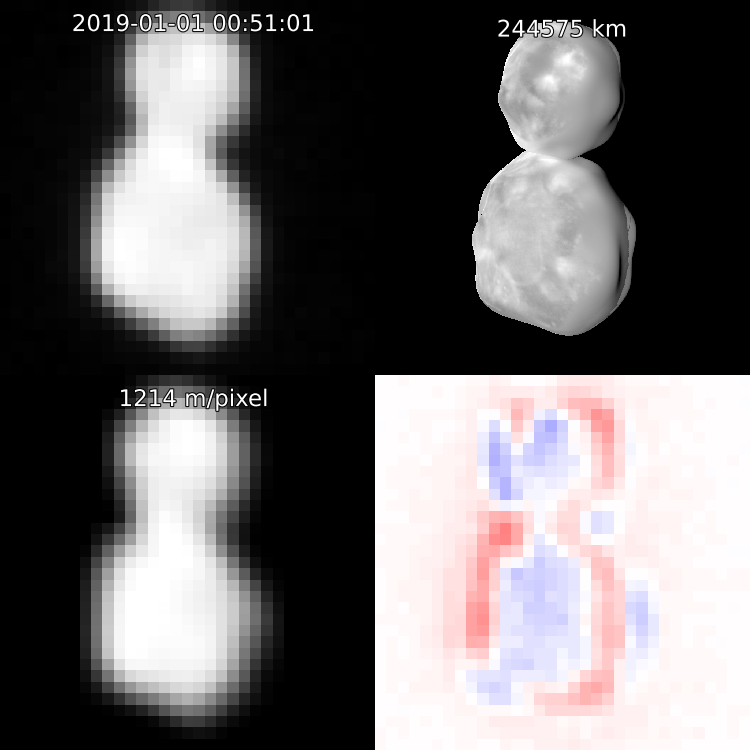}{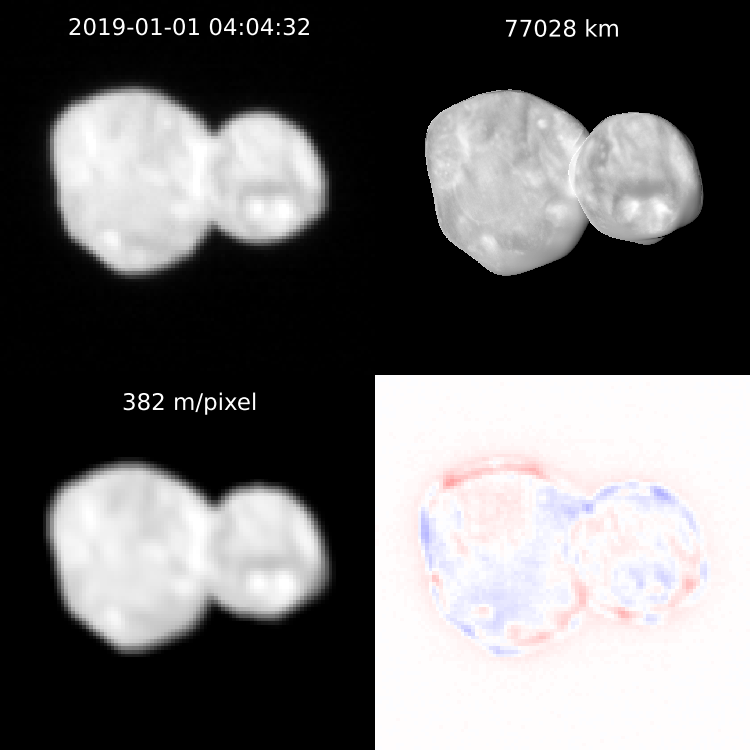}
\caption{Four example distant images sequence stacks
(\texttt{APROTNAV\_L1\_2018365A},\texttt{APROTNAV\_L1\_2018365H},\texttt{APROTNAV\_L1\_2019001C}, and \texttt{APROTNAV\_L1\_2019001K})
and their fit to the model.
For each, the upper left shows the stacked, radiometrically-corrected image sequence obtained by LORRI,
the upper right shows a high resolution render of the shape model with correct orientation and lighting,
the lower left shows the shape model rendered at the resolution of the LORRI image and
convolved with the LORRI PSF, 
and the lower right shows the difference between the images, blue for brighter in the 
data, red for brighter in the synthetic image.
The time of observation at the spacecraft, the distance between Arrokoth and the spacecraft, 
and the resolution of the image in kilometers per pixel are annotated in each image.
Note that these distant and low-resolution images provide the strongest constraint on the shape of the ``neck'' between the two lobes because this area was obscured in the closer images.
\label{fig:quad1}}
\end{figure*}

Fitting the shape of Arrokoth was challenging due to the extremely sparse dataset available.
Arrokoth showed no discernible lightcurve variation in HST/WFC3 photometry \citep{2019Icar..334...11B}, 
which (as we discuss below) is both likely why Arrokoth was discovered in the first place, and shows why the fraction of contact binaries 
in  the Kuiper Belt could be presently underestimated \citep{2021Icar..35614098S}.
This lack of a useful pre-flyby lightcurve meant that the rotational period of Arrokoth could only be constrained by New Horizons
observations.
However, even on approach to Arrokoth, the New Horizons photometric observations also showed no discernible variation in its lightcurve 
\citep{2020Sci...367.3999S}.
This is in contrast to the small satellites of Pluto, which did all show lightcurves on approach to Pluto, allowing their rotational 
periods to be constrained to high precision independently of their shapes or poles \citep{2016Sci...351.0030W,2021psnh.book..457P}.
For Arrokoth, all constraints for shape, pole, and rotation rate would all have to come from the limited number of resolved images.

\subsection{LORRI Image Analysis}

\begin{figure*}
\plottwo{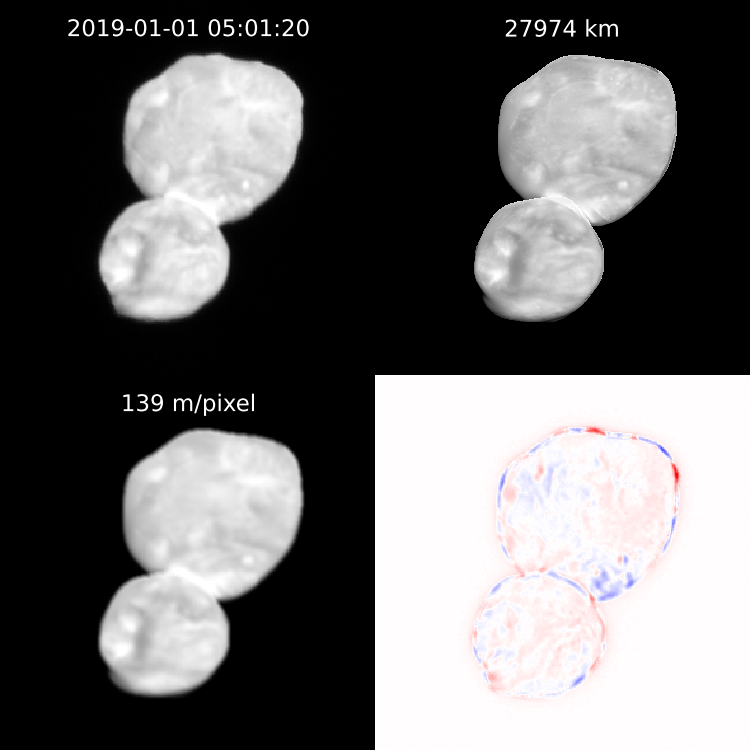}{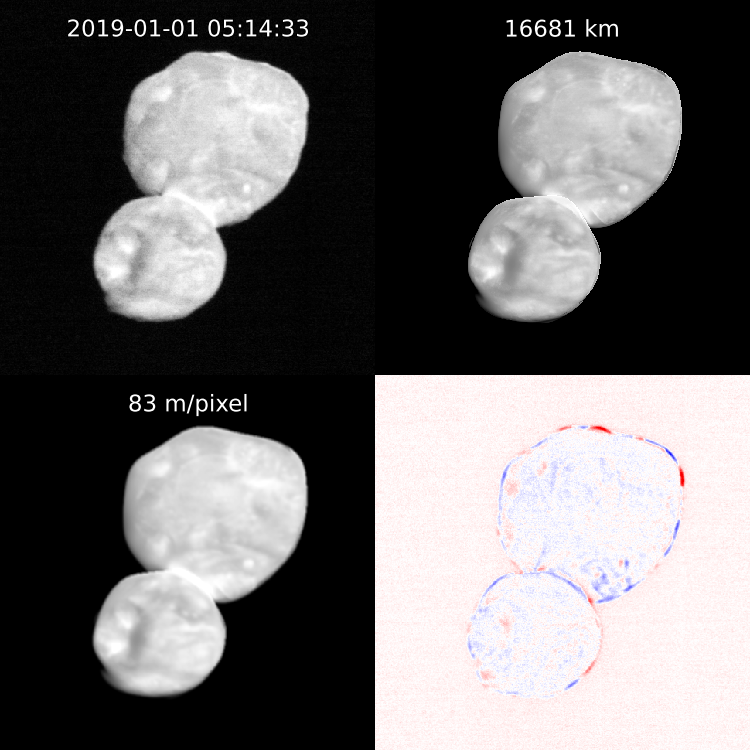}
\plottwo{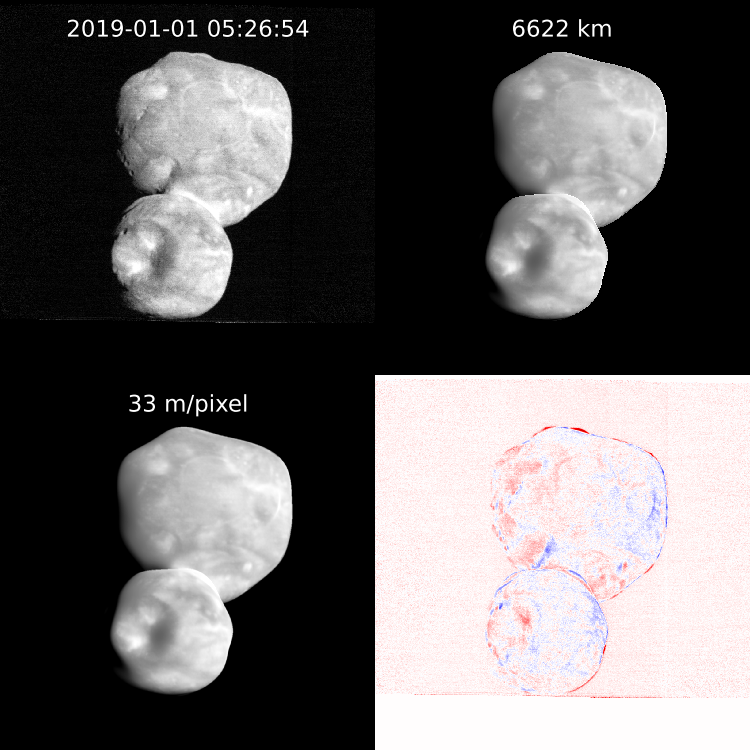}{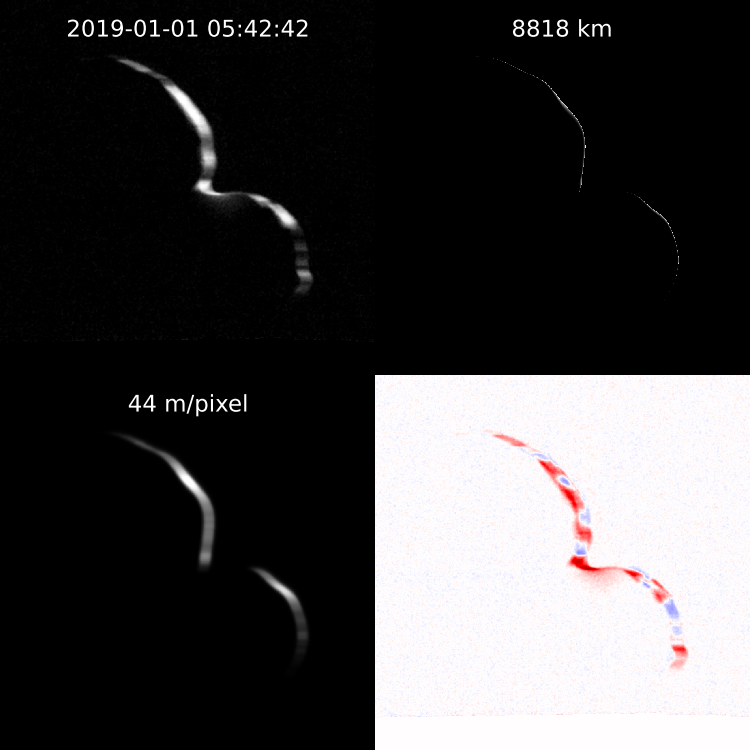}
\caption{The closest four image sequences (\texttt{CA04},\texttt{CA05},\texttt{CA06},\texttt{CA07}),
in the same format as Figure \ref{fig:quad1}.
\texttt{CA04} and \texttt{CA05} are similar view angles, but \texttt{CA05} is higher resolution, while \texttt{CA04} 
was a deeper image sequence with many more images (and so less noisy).
\texttt{CA06} is the highest resolution image, and at considerably different phase angle (see Table \ref{tab:im}).
This meant that much of the fitting process was balancing the fit from \texttt{CA06} with the fit for all the other images.
\texttt{CA07} is the only lookback image, and is considerably smeared (in the horizontal direction) 
by the motion of the spacecraft relative to Arrokoth as the 0.2 second exposures were taken.
\label{fig:quad2}}
\end{figure*}

New Horizons began obtaining images of Arrokoth on August 16, 2018, when the KBO was still much smaller than the 1.04 arcsecond pixels of
LORRI \citep{2020PASP..132c5003W}.
LORRI is a 20.8 cm f/12.6 RC camera with $1024\times1024$ illuminated pixels on an E2V
Technologies Model 47-20 CCD sensor \citep{2008SSRv..140..189C}.
LORRI has no filters, and its spectral response function (0.61 ${\mu}m$ pivot wavelength) is almost entirely determined by the CCD sensor \citep{2008SSRv..140..189C}.
The point spread function (PSF) of New Horizons LORRI is well-documented, and slightly asymmetric \citep{2020PASP..132c5003W}.
Arrokoth first became larger than the LORRI PSF on December 29, 2018, which was the first confirmation that it was not a separated binary
\citep[one possible interpretation of the occultation profile,][]{2020AJ....159..130B}, but a rather single body.
No satellites of Arrokoth were detected in these or any subsequent images during the flyby \citep{2020Sci...367.3999S}.

The earliest images that usefully show the shape of Arrokoth were obtained on December 31, 2018, the day before the flyby.
In total, 30 different resolved image sequences were useful for shape fitting, and they are listed in Table \ref{tab:im}.
Notably, only the southern half of Arrokoth was illuminated and visible in the LORRI images.
The basic image analysis procedure was identical for all of the approach images, with some slight modification for the closest image 
sequence, \texttt{CA06}.
Due to the desire to maximize the Jupiter gravity assist and reach Pluto as fast as possible, the New Horizons 
post-Pluto trajectory points roughly at $l$=16$^\circ$, $b$=-13$^\circ$ in galactic coordinates \citep{2022SSRv..218...11N}.
In practical terms, this means that the spacecraft is heading toward an extremely crowded star field near the galactic center,
which both made the discovery of Arrokoth very challenging \citep{2024PSJ.....5..196B},
and also enabled the stellar occultations that first discovered its contact binary shape \citep{2020AJ....159..130B}.
A side effect of this trajectory is that even the very short exposure images of Arrokoth have detectable background stars.
We were thus able to connect these stars to the Gaia DR3 catalog \citep{2023A&A...674A...1G} and derive the absolute pointing of the images
in the ICRF \citep[International Celestial Reference Frame;][]{1998AJ....116..516M} frame.
By combining this absolute pointing knowledge with the Arrokoth orbit determined from HST and occultation observations \citep{2018AJ....156...20P}
and the trajectory of New Horizons as reconstructed from both optical navigation and radio data \citep{2022SSRv..218...11N},
we can know exactly the ICRF geometry of each observation.
This knowledge is critical for the following shape/pole analysis, and would be much more difficult to obtain without
the Gaia-matched background stars.
The images were radiometrically calibrated using the known LORRI radiometric calibration parameters \citep{2020PASP..132c5003W},
converting them from instrument DN/second to ``I/F'', the ratio of sunlight reflected back to the detector, 
where I/F=1 is full reflectance and I/F=0 is no reflection.
In order to maximize the signal-to-noise ratio (SNR) of the images, one image from each sequence was chosen as the fiducial,
and the rest of the I/F-converted images are reprojected so that Arrokoth was in the same pixel position in all the images.
These reprojected I/F images were then combined using a sigma-clipped median method (sigma-clipped stats in Astropy);
this proved effective at filtering out cosmic rays, which were typically a larger source of noise in the images than CCD read noise.
The resulting stacked images should then provide I/F images for each sequence with known geometry and the highest SNR possible.
The image sequences and their particulars can be found in Table \ref{tab:im}.

\begin{figure}
\plotone{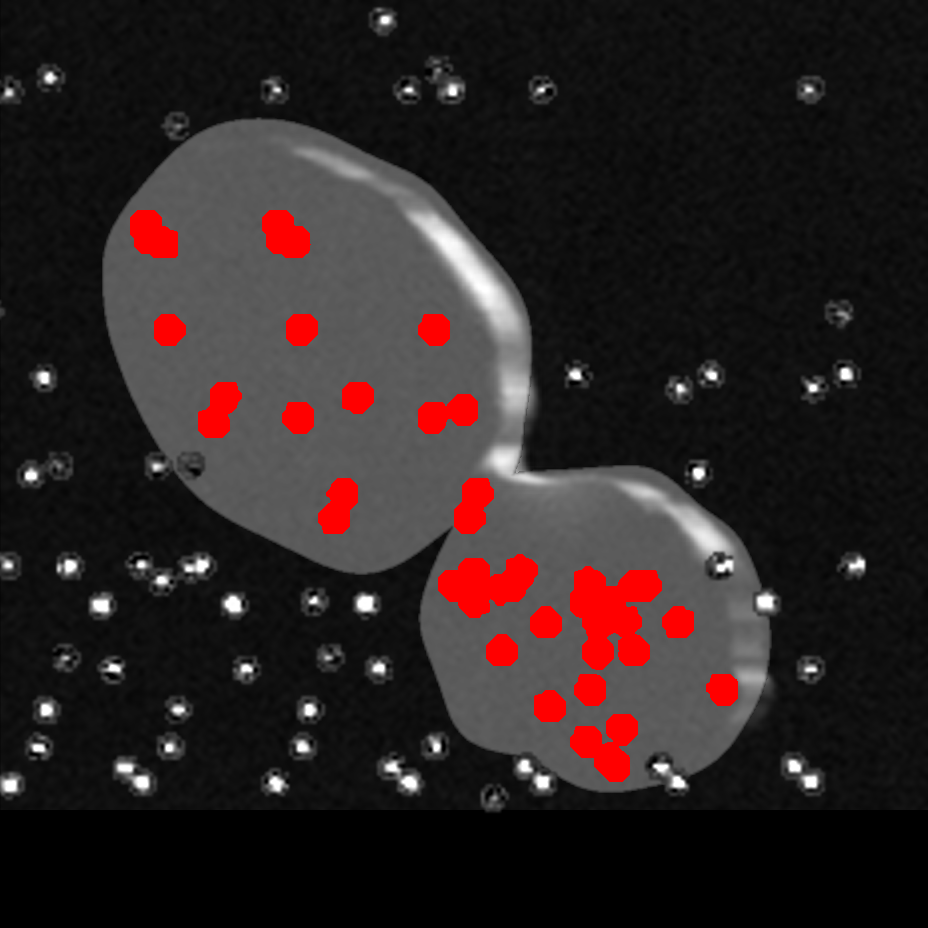}
\caption{The ``lookback'' image sequence \texttt{CA07} superimposed with the shape model, the stars that were occulted by Arrokoth in the sequence in red, and unocculted stars in white. These were the only only 4$\times$4 images used for the analysis, due to the long required exposures. Stars were considered as occulted if their PSF-fit flux dropped more than 50\% for the 0.2 second images. Note that the only stars occulted in the neck region are consistent with the thin neck of the shape model.}
\label{fig:ca07}
\end{figure}

In addition to the imaging before close approach, New Horizons also obtained one useful ``lookback'' sequence (\texttt{CA07}) with the
LORRI camera in 4$\times$4 mode.
This mode bins the illuminated $1024\times1024$ pixels on LORRI's sensor chip four by four, to produce a $256\times256$ image with pixels
4.16 arcseconds across on the sky \citep{2008SSRv..140..189C}.
The 4$\times$4 mode allows for much deeper, longer exposure images than the unbinned 1$\times$1 mode, and the loss of resolution is acceptable, as it is only 
slightly worse than the pointing stability of New Horizons under thrusters 
\citep[New Horizons has no reaction wheels;][]{2008SSRv..140...23F}.
The exposure time of images in 1$\times$1 mode are thus generally kept under the thruster firing frequency (i.e. $\leq$0.15 seconds), 
while the 4$\times$4 mode exposure times are only limited by the onboard software.
The \texttt{CA07} sequence consisted of 336 total images with 0.2006 second exposure times, of which 13 contained the faint double crescent of 
Arrokoth.
Like the 1$\times$1 images, we registered each image to the Gaia DR3 stars and used that to correct the World Coordinate System (WCS) of the images.
While the approach images were looking in the rough direction of the galactic center, the lookback images where looking opposite to 
that, still within the galactic plane and with plenty of stars for registration 
(indeed, substantially more than the 1$\times$1s due to the longer exposure times).
The images were then converted to I/F in the same way as the 1$\times$1s, but using the 4$\times$4 radiometric constants \citep{2020PASP..132c5003W}..
The I/F images were then shifted and stacked to produce the final crescent image (see Figure \ref{fig:quad2}).
Because of the high solar phase angle (averaging 152$^\circ$), the maximum I/F of the lookback image was only 0.0015, compared to
0.127 for the close approach images.
This high-phase image also provided a strong constraint to the overall bidirectional reflectance function fit for the full object.
In addition, Figure \ref{fig:ca07} shows \texttt{CA07} with the shape model superimposed, and the stars that Arrokoth appeared to occult during the scan of the \texttt{CA07} sequence.

\subsection{Shape Fitting Analysis Methods}

\begin{figure}
\plotone{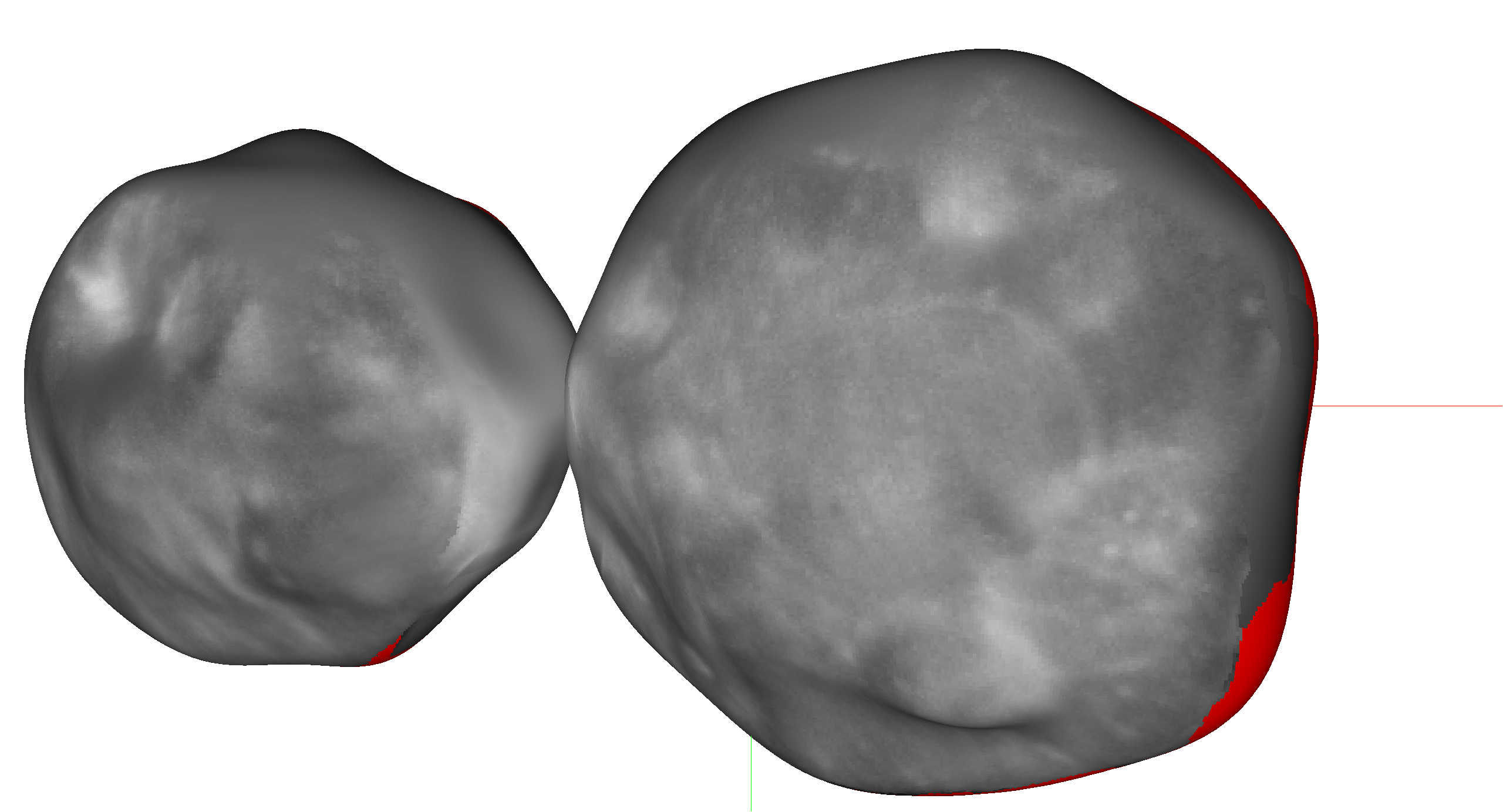}
\caption{The shape model of Arrokoth as viewed from the negative (south) pole. Note the roughly hexagonal shape of Wenu, the larger lobe, 
and the similar (but less well-defined) hexagonal shape for Weeyo, the smaller lobe.
This profile may be due to the formation of Arrokoth from ``mounds'', as detailed in \citep{2023PSJ.....4..176S}.
Note that the bright features in Akasa Linea (between the lobes) are relatively small, 
but appear relatively larger in the close approach images due to the viewing angle.
\label{fig:poleon}}
\end{figure}

To simultaneously fit the shape, pole, and albedo, we used a forward-modeling method of rendering a parametric shape model
and then attempting to find the best fit to the data.
This forward-modeling method was required given the limited amount of data, and enabled a fit for the shape to be detailed on the imaged
half of Arrokoth, while allowing detail on the unimaged half to be minimized.
Forward-modeling that shape with a purely CPU-based approach would be time-consuming, and also unnecessary.
We developed an OpenGL-based forward modeling process for modeling the shapes and poles of the small satellites of Pluto (which have 
even sparser datasets), as documented in \citet{2021psnh.book..457P}.
This same method was used for the initial Arrokoth models published in \citet{2019Sci...364.9771S} and \citet{2020Sci...367.3999S} 
and used for the analysis in \citet{2020Sci...367.6620M}, \citet{2022JGRE..12707068K}, and others.
However, this shape fitting method had some key deficiencies, both in terms of the quality of its output and in the speed of execution of the code.
The New Horizons KEM2 extended mission proposed to improve both the model and fitting technique to produce what would hopefully
be the definitive shape model product for Arrokoth that is presented here.

The shape fitting model used here was effectively a rewrite of the older shape code to allow a transition from the PyOpenGL
library to ModernGL\footnote{https://moderngl.readthedocs.io/en/stable/index.html}.
ModernGL is a currently maintained OpenGL wrapper for Python 3 that implements many of the slower functions in C to maximize execution speed.
The shape fitting code is implemented as a Python script that is typically executed on the Linux command line,
with command line arguments to define behavior.
The shape fitter begins by creating a headless OpenGL context, which is useful to enable remote execution of the code.
The OpenGL buffers for shape model and texture are defined here, and subsequent operations modify them without reallocation.
The GLSL shaders are also defined and compiled here.
There are two sets of shaders, a vertex and fragment shader to calculate the solar illumination on surface, and
a vertex and fragment shader that calculate a shadow map, which is rendered directly to a texture.
Next, a plain-text input file is read, which defines the rotational properties of the object, the parameterized shape,
the photometric properties of the object (including if an albedo map is being used),
and finally the input I/F images to fit.
For Arrokoth, the overall shape was defined by two parametric shape models (one per lobe) and their offsets from center,
though the code can use a single parameterized shape model as well.
The lobes are parameterized with an ``Octantoid'' formalism as described by \citet{2012A&A...543A..97K}.
This is similar to a spherical harmonic model, but has the advantage that the lowest-order parameters precisely define a triaxial
ellipsoid.
If an albedo map is defined, it is loaded in here; it is stored as floating point array in a FITS file.
The map projection is also defined in the input file, which for Arrokoth was two south polar azimuthal equidistant maps 
(like the United Nations flag, but centered on the south pole), with a separate projection for each of the two lobes.
This projection was chosen because it had the least amount of distortion for the high resolution images, 
and two separate projections were used to eliminate any ambiguities between the lobes.
The script has predefined OpenGL UV maps for the supported map projections.
The photometric parameters for the model are defined at this point too.
This includes a global albedo, which is multiplied by the albedo map if a map is used.
Each of the images are defined with the input I/F FITS file (as produced with the procedure above), a number to define image type,
and pixel x/y image offsets.
Each of the I/F FITS files contain the observation time in SPICE Ephemeris Time (ET) at the spacecraft,
the WCS-format pointing information, and the image data itself.
The ET and WCS are used to create static OpenGL rotation and view matrices for each image.
The image ``mode'' is used to define the PSF used for the image, with most of the images just using the native LORRI PSF,
and with the \texttt{CA07} images using a smeared LORRI PSF due to the spacecraft motion.
Each image as designated a fixed weight based on the square root of the number of images used in the stack (proportional to signal to noise) multiplied by the 
spacecraft-target distance (proportional to the angular size of Arrokoth).
This weighting makes it so the $\chi^2$ for all the images is similar for a good solution and 
is required to ensure that the closest images don't entirely swamp the more distant images that probe other parts of the shape.

\begin{figure*}
\plotone{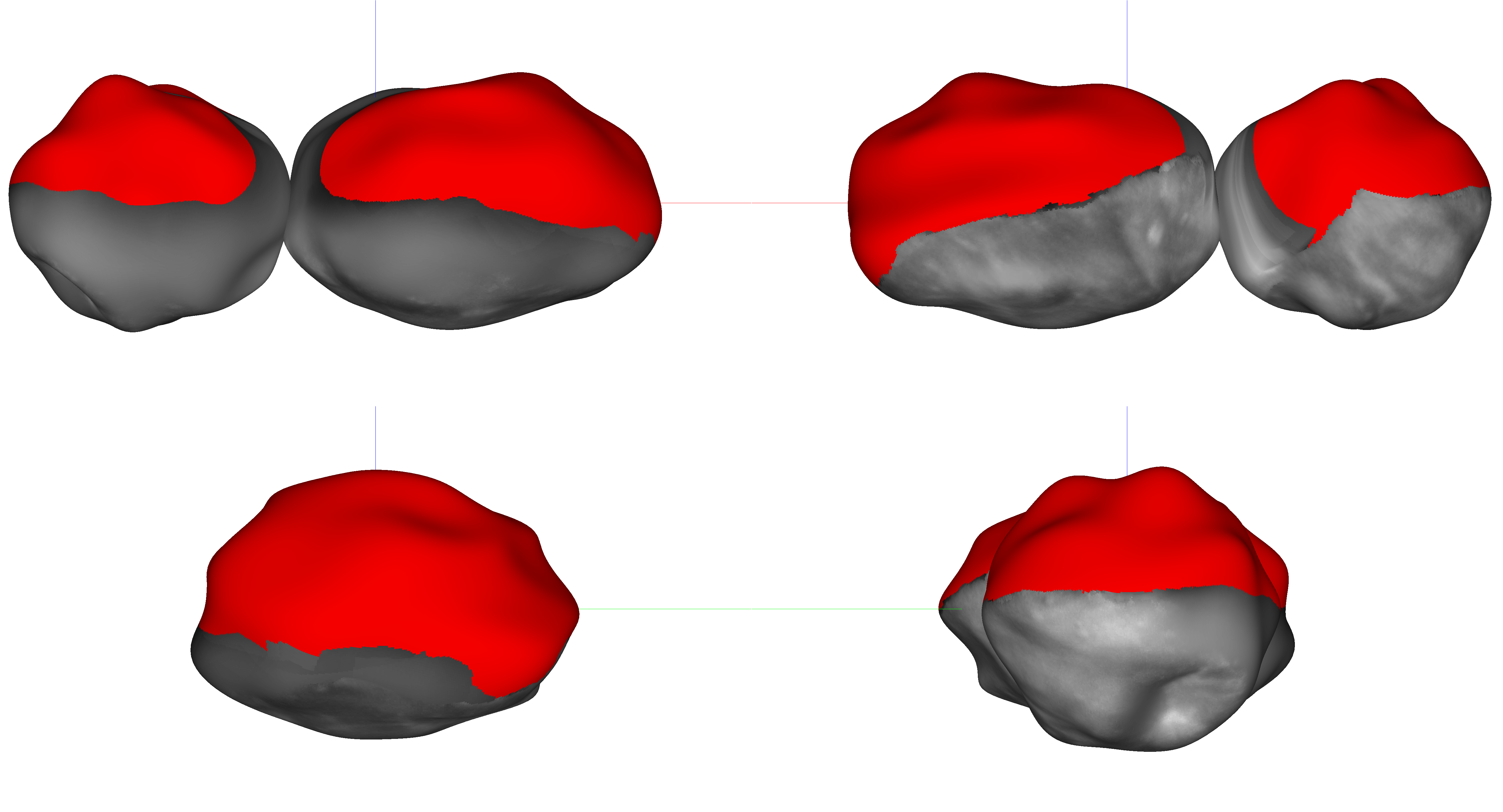}
\caption{The shape of Arrokoth as viewed from its equator. 
The red areas indicate regions that were not well-imaged by New Horizons at closest approach 
(i.e. those areas images at too low resolutions for the albedo fitter). 
The detailed shape of these areas is effectively the reflections of the parametric shapes for the well-imaged parts (due to the nature of the octantoid spherical harmonics), 
though the overall dimensions are well-constrained by the low resolution images (see Figure \ref{fig:quad1}).
\label{fig:eq1}}
\end{figure*}

With the input file read and processed, the script then performs the operation defined on the command line.
The most central operation is to calculate the $\chi^2$ residuals for an input file or parameters.
This starts with converting the parametric shape model from parameters to two closed, overlapping meshes. 
The evaluation of the spherical harmonics used the fast ``lpmv'' 
(associated Legendre function of integer order and real degree) function from 
SciPy \citep{2020NatMe..17..261V} using the CPU.
The vertex normals are then calculated and used to define normals for each of the triangular facets.
This facet normal calculation was one of the slowest parts of the older code, and in the new version is performed in GPU
using the PyOpenCL wrapper for OpenCL \citep{2009arXiv0911.3456K}, allowing for considerable speedup.
The updated vertices and normals are then copied to the appropriate OpenGL framebuffers.
The photometric parameters are also copied over to the correct GPU variables for the shader to use.
If the complexity flag was called in the input file, then the complexity factor is calculated here as the sum
of the squares of all the shape parameters higher than first-order.
If the balance flag was called in the input file, the volumes of the lobes are calculated and the offset of center of mass
from the nominal center point is calculated.
The script then integrates through each of the images.
First, the object rotation matrix is calculated from the pole and rotation period, and this rotation matrix is applied to the pre-computed
view matrix and the appropriate frame buffers updated.
The synthetic image is then rendered with OpenGL.
The GLSL shader uses the geometry provided to OpenGL to calculate the surface brightness with a bidirectional reflectance model 
(BDRM) of \citet{2012tres.book.....H}.
The shader uses the global photometric parameters and the albedo from the texture map to calculate the effective I/F
for each pixel of the image.
To define the functional form of the BDRM, we first used
Equations 13 and 14 of \citet{1981JGR....86.3039H} to define a backscatter function:            
\begin{equation}
B = ( 1 - \frac{tan|g|}{2h}(3-exp(-\frac{h}{tan|g|}))(1-exp(-\frac{h}{tan|g|}))) exp(-\frac{w^2}{2})
\end{equation}
where $g$ is the Sun-Target-Observer phase angle in radians, 
$h$ is a parameter that defines particle spacing and was -7.05$\times10^{-6}$ in the best fit, 
and $w$ is single-scattering albedo (as defined by the albedo map).
Using Equation 14 of \citet{1995Icar..113..134M}, we define a Double Henyey Greenstein function for the particle phase function as:
\begin{equation}
p_{HG} = \frac{1+c}{2}\frac{1-b^2}{(1 - 2b~cos~g + b^2)^{3/2}} + \frac{1-c}{2}\frac{1-b^2}{(1+2b~cos~g+b^2)^{3/2}}
\end{equation}
where $c$ and $b$ are unitless parameters that define the shape of the phase function;
$c$ was 0.564
and 
$b$ was 0.356 in the best fit.
We also used the Ambartsumian-Chandrasekar H function approximation from \citet{2012tres.book.....H} Equations 8.28 and 8.56:
\begin{equation}
r_0 = \frac{1-\sqrt{1-w}}{1-\sqrt{1+w}}
\end{equation}
\begin{equation}
H(x) \simeq 1/(1 - w x (r_0 + \frac{1}{2}(1-2r_0x)ln(\frac{1+x}{x})).
\end{equation}
Note this relation was incorrectly listed in the first edition of that book \citep{1993tres.book.....H}, 
and the version in \citet{2012tres.book.....H} should be used.
This is finally combined for a total BRDM function by modifying Equation 16 of \citet{1981JGR....86.3039H} by dropping 4$\pi$ to convert from irradiance to radiance and multiplying $H(\mu_{Sun})$ by $\mu_{sun}$:
\begin{equation}
r = \frac{ \mu_{Sun}w }{ \mu_{Sun} + \mu_{obs} } ((1+B)p_{HG} + \mu_{sun}H(\mu_{Sun})H(\mu_{obs}) - 1)
\end{equation}
where $\mu_{Sun}$ is the cosine of the angle between the surface normal and the direction of the Sun,
and $\mu_{obs}$ is cosine of the angle between the surface normal and the direction of the observer.
These functions were all directly implemented in GLSL, allowing them to be executed at speed in the GPU.

The rendered image is then copied from the GPU framebuffer back to CPU memory.
Next, the image is convolved with the appropriate PSF.
By default, this is done in CPU, but can be optionally performed by GPU using CUDA through Numba \citep{2015llvm.confE...1L}.
The CUDA-based convolution generally executes much faster than performing the convolution in CPU, even with extra memory copies to and from the GPU.
Finally, the rendered image is subtracted from the actual image, the difference image squared and summed, and multiplied by
a weight based on the angular size of the object.
This per-visit $\chi^2$ is calculated and summed for all the image stacks, and the complexity and balance factors added if requested, and the result
returned as the effective $\chi^2$ for optimization.

\begin{deluxetable*}{cccc}
\tablecaption{Parameters of the best-fit shape solution. 
\label{tab:param}}
\tablehead{ & \colhead{Full Body} & \colhead{Wenu} & \colhead{Weeyo} }
\startdata
a Length &   34.546 km & 20.139 km & 15.041 km \\
b Length &   19.838 km & 19.838 km & 14.378 km \\
c Length &   13.822 km & 13.726 km & 13.625 km \\
a/b Ratio & 1.741 & 1.015 & 1.046 \\
a/c Ratio & 2.499 & 1.467 & 1.104 \\
Diameter of a Sphere of Equal Volume: & 19.896 km & 17.349 km & 13.845 km \\
Rotational Period & 15.4453 hours & & \\
Rotational Pole Right Ascension & 319.37$^\circ$ & & \\
Rotational Pole Declination & -25.588$^\circ$ & & \\
Pole inclination to heliocentric orbit & 100.39$^\circ$ & & \\
Pole inclination to approach direction & 41.096$^\circ$ & & \\
\enddata
\tablenotetext{}{
The ``a'' direction points from Weeyo to Wenu,
the ``c'' direction points to the rotational pole, 
and ``b'' is mutually perpendicular.
See text for a discussion of uncertainties.
}
\end{deluxetable*}

The shape, pole, photometric properties, and individual image offsets can be allowed to vary for the fit.
In practice, many iterations were performed with just one set of those parameters allowed to vary;
a global parameter mask was set on the command line and the optimization functions only performed on the parameters in the mask.
The available optimization functions included the SciPy implementations of the Powell \citep{powell1964efficient} and Nelder-Mead \citep{nelder1965simplex}
function minimizers.
The former optimizes parameters individually and is thus preferable for the shape and offsets, 
while the latter optimizes parameters all at once and is thus preferable for the highly covariant pole and photometric properties.
In addition, two special optimizers were used, 
one of which only optimized the offsets individually,
and another which only optimized the shape parameters one by one in order of absolute magnitude.
This enabled a quick optimization that focused first on the parameters most relevant to the solution and could be cancelled
before spending time on the less important parameters.
It was possible to capture incomplete solutions because the optimizer would output the current state after optimizing each parameter.
This could then be fed back into the command line using a flag that looked for parameters in STDIN,
and could plot or save those parameters into an input file.
The models could be visually inspected with a separate script that transformed the parametric model
into a mesh and displayed it in an interactive way using Vispy \citep{2022zndo...6795163C};
this script was also used to generate Figures \ref{fig:poleon} and \ref{fig:eq1}.

A low-order model (essentially two triaxial ellipsoids) was initially used, and then more orders were gradually added on to increase the 
complexity, with the final model using order-9 octantoids.
Once a good initial model with uniform albedo converged, it was used to fit an albedo map, and then refit with the albedo map, the map refit, and so on.
The albedo map was produced by first reprojecting the 30 approach images using the shape model, saving both the I/F brightness and the angles needed for the bidirectional reflectance model (BDRM). 
A best-fit was then found for the albedo map using the reprojected images, weighted for the original resolution of the reprojected image.
This method allowed us to combine all the images, both the high resolution close approach images that showed detail on the surface 
and the more distant images that showed other aspects of the shape, to be combined into one albedo map.
Fitting the shape and albedo map was a highly iterative process, which took nearly a year to finally converge on a best fit shape.
The initial model developed after flyby had an incorrect pole and did not model the variability in surface albedo, 
resulting a much flatter shape, and which had no constraints on the unlit side of Arrokoth.
This model was published in \citet{2020Sci...367.3999S} and in the original PDS delivery;
testing the model from \citet{2020Sci...367.3999S} showed that its $\chi^2$ was 60\% higher 
than the solution presented below on the same dataset.
The current model corrects the pole and recenters the shape, to be roughly balanced in volume and mass (so that the rotation pole is through the center of volume).
As noted below, this has a significant impact on the interpretation of the model, and is why the solution presented here is 
substantially larger in the c axis (and thus larger volume) than the older model in \citet{2020Sci...367.3999S}.

\section{Shape Fitting Results}

\begin{figure*}
\plottwo{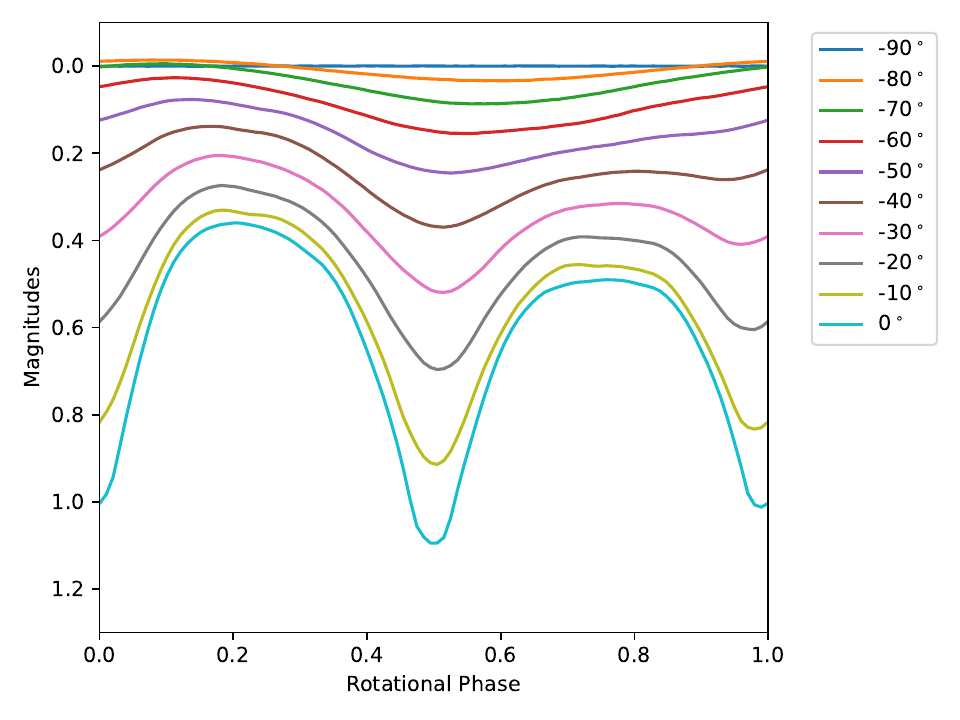}{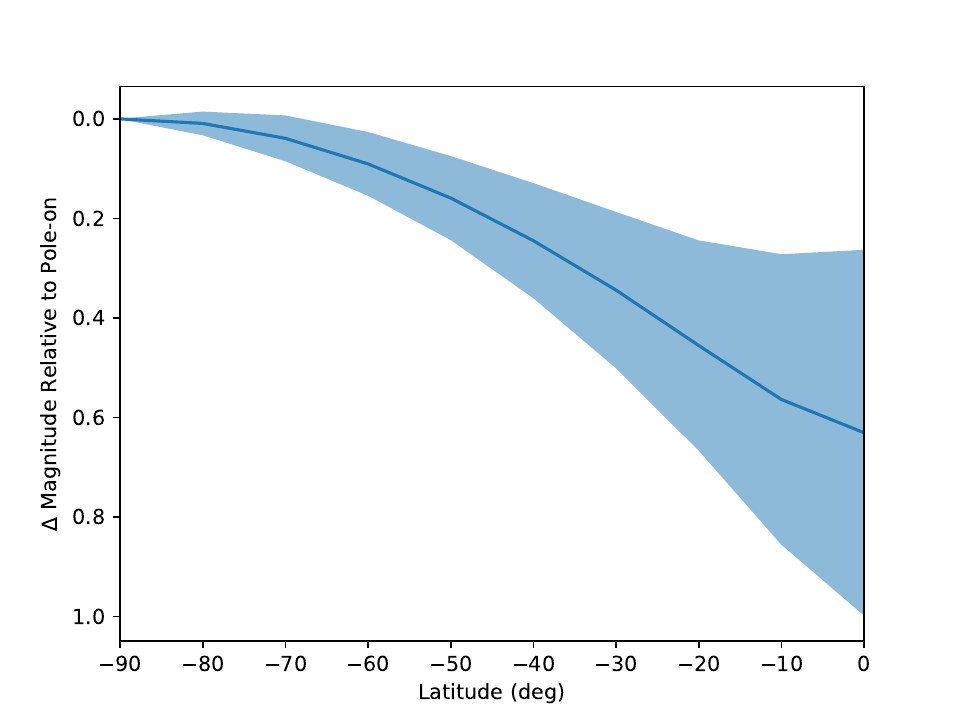}
\caption{The rotation lightcurve of Arrokoth at a Sun-Target-Observer angle of zero, as viewed from various sub-observer latitudes.
The left plot shows the variation in the shape and reflectance of the lightcurve at different latitudes, 
while the right plot shows the variation of the mean reflectance (solid line) and lightcurve amplitude (shaded area) as a function of sub-observer
latitude.
Note that even with uncertainy as to the precise topography of the unlit side of Arrokoth,
the flattened shape of the two lobes of Arrokoth mean that the low latitudes (which show any amount of 
lightcurve variation) are also more than a half magnitude fainter on average than the mean reflectance when viewed 
pole-on.
This result is significantly different from that of \citet{2021Icar..35614098S}, and could substantially affect the statistics of discovered KBOs.
\label{fig:lc}}
\end{figure*}

Comparison plots between the LORRI data and the model can be seen in Figures \ref{fig:quad1} and \ref{fig:quad2},
orthographic projections of the shape model are shown in Figures \ref{fig:poleon} and \ref{fig:eq1},
and the parameters of its shape and pole listed in Table \ref{tab:param}.
The model itself is included in the supplementary material as an OBJ file.
Note that since not all of the shape was seen by New Horizons, the red areas are used in Figures  \ref{fig:poleon} and \ref{fig:eq1}
to denote areas that were not imaged by New Horizons and only constrained by their absence.
The overall shape is bilobate, with the small lobe Weeyo being roughly spherical,
and the large lobe Wenu being significantly oblate.
This is a notable change from the much flatter shape presented in \citet{2020Sci...367.3999S}, 
and largely due to the improved pole solution.
Neither lobe is particularly prolate, although Weeyo is slightly more prolate, and its long axis is aligned
to point at Wenu.
Weeyo contacts Wenu along the equator of Wenu, though the exact contact point is obscured in the 
high resolution images (see Figure \ref{fig:quad1} for the best image constraint on the neck).
The apparent volume ratio between Wenu and Weeyo is 1.967:1;
though the precise value could vary based on the unseen areas on the unlit side of Arrokoth, it cannot vary too much
from 2:1 without being invalidated by the approach images.
The overall body is much more elongated and flattened than the individual lobes, and this is reflected its rotation
lightcurve shown in Figure \ref{fig:lc}.

Both lobes appear to be roughly hexagonal in shape along their equators.
Hexagonal shapes are difficult to produce with octantoid spherical harmonics, 
so this is not likely to be an artifact. 
Weeyo's long axis is aligned to its contact point with Wenu, 
while Wenu appears somewhat misaligned as viewed from the pole (see Figure \ref{fig:poleon}).
This contact region, named Akasa Linea, appears to be filled with high albedo material \citep{2021Icar..35613723H}, which obscures
the shape information and makes fitting the shape of the contact region difficult.
Overall the nominal albedo of the body (using the BDRF described above) varies from $\sim$0.017 to $\sim$0.08,
with the bright albedo features found in discrete areas on both lobes, notably inside Sky Crater on Weeyo,
Akasa, and several areas of Wenu, including the circular Ka'an Arcus feature in the center of Wenu.
Ka'an Arcus appears to encircle an area of elevated terrain, consistent with it being a separate ``mound''
\citep[as proposed by][see Figure \ref{fig:eq1}]{2023PSJ.....4..176S}.
The overall profile of Wenu is more similar to cylindrical prism than a perfect triaxial ellipsoid,
with near-vertical sides topped by a flat surface with a raised central mound (see Figure \ref{fig:eq1}).
In contrast, Weeyo appears to be close to a spherical object with significant surface terrain dominated by Sky Crater.

Figure \ref{fig:ca07} shows the lookback \texttt{CA07} sequence as compared to the background stars and shape model,
using a similar analysis to \citet{2020Sci...367.3999S}.
The occulted stars are generally consistent with the shape model, though there are a few stars on the unlit limb of Arrokoth where it is hard to tell if they show more detail in the shape, or were just partially occulted.
Notably, the thin neck of the shape model is consistent with the stellar occultations.

The fitting methods used here were not perfect, and they did not fully fit the shape and albedo to the limits of the data.
A notable deficiency is the area of Wenu closest to Akasa which hosts a small bright albedo feature.
This area never quite converged, and the residuals of this poor fit stand out in Figure \ref{fig:quad2}.
Also, the full shape of Sky Crater is below the resolution of the parametric model, particularly along the rim.
This is also true of all of the small craters that are visible near the terminator.
In general, however, the overall shape of Arrokoth was well fit by the model.
Future work is planned to use this low-frequency model as a basis for higher resolution stereophotoclinometry of the
well-imaged parts of the surface.

The model presented here is a best-fit, and considerably more analysis is required to produce uncertainties for the dimensions and shape of Arrokoth.
Table \ref{tab:param} presents the dimensions of the overall body and lobes to a precision of 1 meter, 
but the actual uncertainty is larger.
The uncertainty for the $a$ and $b$ axes are likely better than 100 meters, as they are constrained from multiple viewing angles; see Figure \ref{fig:poleon}.
However, as the red areas in Figure \ref{fig:eq1} show, there is much more uncertainty in the $c$ axis
since only half of Arrokoth was seen at high resolution.
The uncertainty in the $c$ axis could thus be as high as 1 km due to local topography in the northern hemispheres of the lobes, but not much more given the constraints from the CA07 occultation (Figure \ref{fig:ca07}).

\section{Discussion}

\subsection{On the Frequency of Contact Binaries in the Kuiper Belt}

The synthetic rotation lightcurve of Arrokoth projected at various sub-observer latitudes is shown on the left in Figure \ref{fig:lc}, 
and the amount of lightcurve variation, or amplitude,  as a function of sub-observer latitude is shown on the right in Figure 
\ref{fig:lc}.
These figures were created with the same forward-modeling code as described above, but with a fictional observer looking at
the shape at precisely solar opposition (i.e. Sun-Target-Observer angle of zero), and then taking the sum of the rendered image pixels
to produce an effective reflectance for each point in the lightcurve.
This is done for 11 different longitudes to produce the effective zero-phase lightcurves shown in Figure \ref{fig:lc}.
Therefore, the amplitudes of the rotation lightcurves of an Arrokoth-shaped object can vary greatly, depending on the sub-observer latitude.
Notably, seen from Earth, the sub-observer latitude on Arrokoth is closest to the -80$^\circ$ line, thus showing why no rotation lightcurve amplitude for it was found in HST photometry
\citep{2019Icar..334...11B}.

The discovery of Arrokoth by HST was enabled by its low-amplitude rotation lightcurve, as it was just barely visible in HST's WFC3 
\citep[roughly 144 photons per 350 second image;][]{2019Icar..334...11B}.
%, enabling it to be consistently recovered by HST.
In general, most known KBOs do not have lightcurves with large amplitudes, and as succeeding surveys try to probe deeper, they generally only report objects that have been
consistently recovered at close to the minimum detectable brightness for the survey \citep[e.g.][]{2024PSJ.....5..227F}.
This situation is compounded by the fact that objects with a consistent brightness are much more likely to have the required multiple-opposition follow up observations, and thus the well-determined orbits that are required to enable targeted lightcurve follow-up.
A KBO with a high amplitude lightcurve is much less likely to be recovered in every observation in a survey program,
since traditional surveys only try to recover their previous discoveries on one or two nights.
That effect is especially pronounced for KBOs that are just bright enough to be detected at the peaks of their brightness, 
but are undetectable in the same survey if their brightness drops by half a magnitude (as in some of the curves in Figure \ref{fig:lc}).
This has likely led to a considerable bias in the documented KBO rotation lightcurves towards objects that do not reveal any shape information.
\citet{2021Icar..35614098S} recognized this bias shortly after the Arrokoth flyby, but their analysis was based on the idea that Arrokoth's shape was approximated by two spheres in contact with each other.
As Figure \ref{fig:lc} shows, this effect is even more pronounced for the actual flattened contact binary shape of Arrokoth.
Lightcurve-based estimates of contact binary frequency 
\citep[e.g.][]{2019AJ....157..228T}
therefore may be considerably underestimating the frequency of contact binaries in the Kuiper Belt.
The Vera Rubin Observatory offers a potential antidote to this bias,
as its wide and deep ecliptic plane survey 
\citep{2023ApJS..266...22S} 
may be able recover faint KBOs of variable brightness, 
and then offer a regular cadence of observations to detect that lightcurve variation.
Arrokoth is a very typical Cold Classical KBO in its orbital characteristics \citep{2018AJ....156...20P}, and future KBO lightcurve studies should seek
to test if it is typical in shape as well.
\citet{2022PSJ.....3...23P} presents several KBOs that were observed by New Horizons at high solar phase angles and are good candidates to perform this test.

\subsection{Implications for Formation}

The shapes of the two lobes of Arrokoth are particularly interesting because they are very likely to have formed separately and later come together
to form the contact binary shape \citep{2020Sci...367.6620M}.
As noted in the introduction, the dominant theory for the formation of KBOs is through the SI, which can rapidly collapse dust from the circumsolar
disk into solid bodies \citep{2017ApJ...839...16C}, many of them initially in loosely bound binary orbits \citep{2019NatAs...3..808N}.
\citet{2023PSJ.....4..176S} have shown that this process can accrete objects in such a way that results in bodies seeming to form ``mounds'',
and the shape model presented here is fully consistent with that process.
\citet{2020Sci...367.6620M} proposed that the two components of Arrokoth were initially on a bound orbit and came together under gas drag,
allowing their mutual orbit semi-major axes to gradually decay until they reached a soft contact.
Nothing in the shape model presented here is inconsistent with that proposal, including the relatively small contact point between
the lobes that suggests a very gentle process for bringing them together.
Other authors have suggested that the lobes of Arrokoth came together as a result of later orbital evolution due to impacts or
other interactions with other KBOs \citep{2025NatAs...9...75C}.
The shape of Arrokoth does not necessarily exclude this, but any such orbital evolution would need to be slow enough to 
preserve the shapes of the lobes on contact.

The shape model presented here does present some new questions and constraints for formation models.
The hexagonal shape of both lobes is extremely curious and is not completely explained by the ``Mounds Model'' \citep{2023PSJ.....4..176S}.
Both lobes being roughly similar size on their shortest axis is also intriguing, and may point to them forming out of
a more disk-like structure of that thickness, but such a conclusion requires future modeling.
The raised central mound on Wenu likely also points to a feature of formation, possibly that Wenu originally formed 
as a more spherical object before accreting the material on its equator.
But again, that possibility would require future formation modeling to fully understand the connection between the shape and formation process.
The presented shape model also provides a powerful test for any alternative KBO formation models to the SI.
The complex shape of Arrokoth being so well explained by the SI means that any alternative theories would need to be able to
reproduce Arrokoth-like shapes as least as well as the SI.

\section{Conclusions}

The updated shape model of Arrokoth provides a much better basis for understanding this small distant world than previous shape models.
The updated, thicker shape model is more consistent with formation though the Streaming Instability \citep{2019NatAs...3..808N,2023PSJ.....4..176S}, 
while still being consistent with proposed models for how the two components of Arrokoth came together \citep{2020Sci...367.6620M}.
The volume ratio of almost precisely 2:1 between the two lobes is intriguing and should provide interesting constraints for future
formation modeling efforts.
The updated Arrokoth shape model implies a significant decrease in reflectance as seen from the equator,
heightening the evidence from \citet{2021Icar..35614098S}
that the contact binary population in the Kuiper Belt may be significantly underestimated.
Arrokoth is fascinating both because it is unlike any other object ever visited by a spacecraft and because it
could be an archetype for thousands of KBOs yet to be explored.

\begin{acknowledgments}
This work was supported by the New Horizons KEM2 Extended Mission.
\end{acknowledgments}

\software{
    Astropy \citep{2022ApJ...935..167A},
    Pyopencl \citep{2009arXiv0911.3456K},
    Scipy \citep{2020NatMe..17..261V},
    Spiceypy \citep{2021zndo...4883901A},
    Numba \citep{2015llvm.confE...1L},
    Vispy \citep{2022zndo...6795163C}
}

\bibliography{refs}{}
\bibliographystyle{aasjournal}

\end{document}